\documentclass[12pt]{iopart}
\usepackage{graphicx}
\usepackage{epsfig}

\begin{document}

\title[]{Direct Photons at RHIC}

\author{Saskia Mioduszewski
\footnote[3]{saskia@bnl.gov} for the PHENIX Collaboration
}

\address{Physics Department, Brookhaven National Laboratory, 
Upton, NY 11973, US}

\begin{abstract}
The PHENIX experiment has measured direct photons in 
$\sqrt{s_{NN}} = 200$~GeV Au+Au
collisions and p+p collisions. The fraction of photons due to direct
production in Au+Au collisions 
is shown as a function of $p_T$ and centrality.  This
measurement is compared with expectation from pQCD calculations.
Other possible sources of direct photons are discussed.
\end{abstract}




\section{Introduction}

One of the most direct probes of the QGP is the radiation of thermal
photons, emerging directly from the thermalized system and providing, in
principle, the temperature in its hottest phases~\cite{shuryak}. 
However, photons are
largely produced in hadronic decays, as well as in 
QCD processes, such as annihilation and
Compton, that occur before thermalization.  Thermally produced
photons must be distinguished from those produced primordially in QCD
interactions and those from final state decays of hadrons. Thus
it is necessary to first determine the yield of photons that are produced in
excess of the photons from known hadronic decay sources (the background).

\section{Background}

Direct photons are extracted from an inclusive photon measurement,
largely composed of photons 
from hadronic decays, such as $\pi^0$, $\eta$, $\omega$, and $\eta^{'}$. 
The dominant source of background 
photons is the $\pi^0$ decay, thus it is essential
to measure the $\pi^0$ yields, with precision, as a function of $p_T$.
Based on the measured $\pi^0$ $p_T$ spectrum, one can simulate 
the spectrum of photons from all known hadronic decays.
This can then be subtracted from the inclusive photons
to obtain a direct photon spectrum.

\begin{figure}[hbt]
\begin{center}
\mbox{\epsfxsize=6in\epsfbox{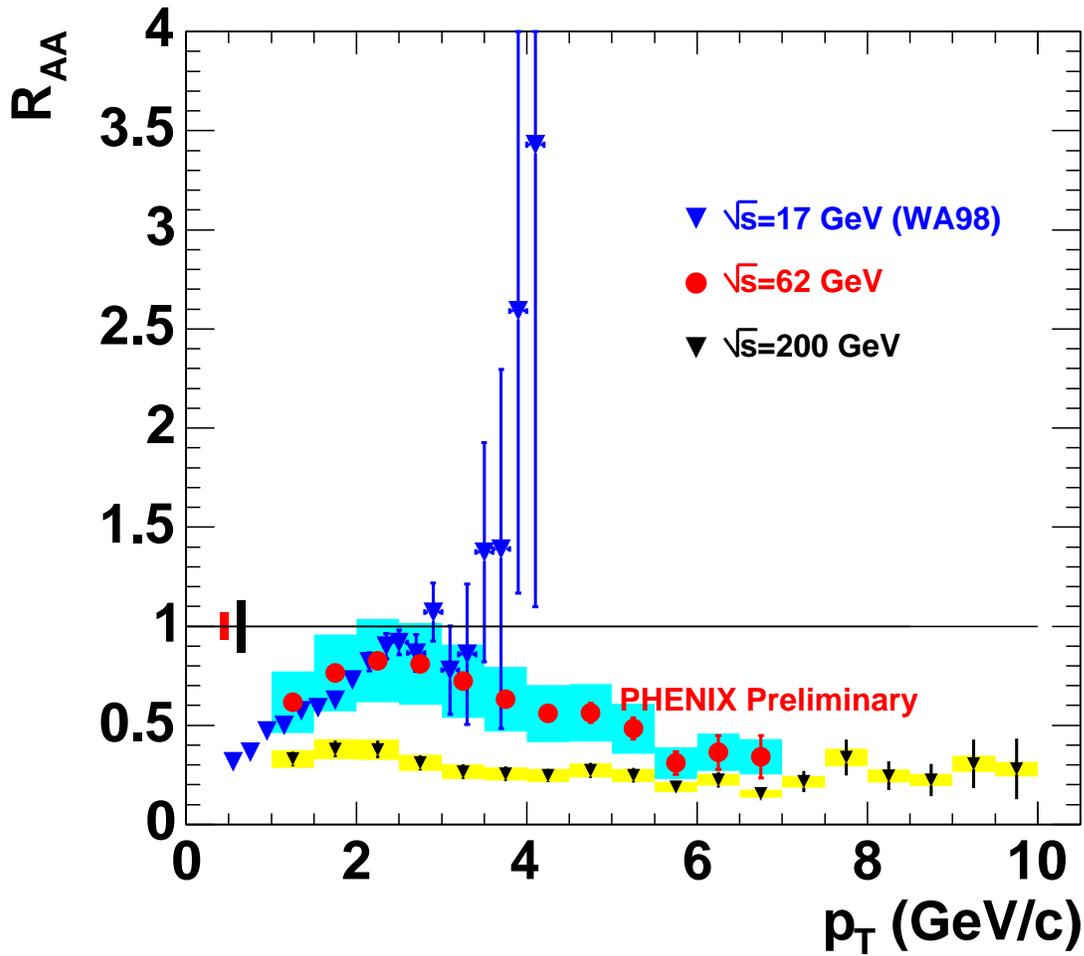}}
\end{center}
\caption{\label{fig:RAA}
$R_{AA}$ for neutral pions in central Au+Au collisions at RHIC at $\sqrt{s_{NN}} = 200$~GeV~\cite{PHENIX_pi0} and $\sqrt{s_{NN}} = 62$~GeV (PHENIX preliminary), and central Pb+Pb collisions at the SPS~\cite{WA98_pi0,David} ($\sqrt{s_{NN}} = 17$~GeV).  The boxes surrounding the data points indicate the
systematic uncertainties that are correlated in $p_T$. 
The black line at 1 shows the percent normalization 
error on the data for $\sqrt{s_{NN}} = 200$~GeV, and the smaller red line for
$\sqrt{s_{NN}} = 62$~GeV.  The normalization error is smaller for 
$\sqrt{s_{NN}} = 62$~GeV because more of the uncertainties are not
simple normalization errors and are thus included in the
systematic error boxes around the points.}   
\end{figure}

At RHIC, we are in the fortunate position of having a large
suppression of $\pi^0$, thus reducing the background photons.
Figure~\ref{fig:RAA} shows the nuclear modification factor for $\pi^0$ at
$\sqrt{s_{NN}} = 200$~\cite{PHENIX_pi0}, 62, and 17 GeV~\cite{WA98_pi0} measured in central A+A collisions.  
At $\sqrt{s_{NN}}=200$~GeV, the suppression of 
$\pi^0$ for $p_T > 3$~GeV/c is a constant factor of $\sim 4-5$, 
relative to expectation from the binary-scaled (scaled by $N_{coll}$)
yields measured in p+p collisions. 
With a large suppression of the hadronic decay background, and 
assuming that the photons are not similarly suppressed due to an 
initial-state effect,
the measurement of direct photons at high $p_T$ should be less
difficult than at lower $\sqrt{s_{NN}}$.

\section{Sources of Direct Photons}

Even after subtracting the photons from hadronic decays, extracting the thermal
photons is still more difficult.  One has to disentangle different sources
of direct photons.  These include pQCD (``prompt'') photons from processes such
as Compton ($q+g \rightarrow q+g$) and annihilation ($q+\overline{q} \rightarrow g+g$), as well as bremsstrahlung (which may also may be modified by the 
medium~\cite{jamal}).
In addition, there is another possible mechanism for producing photons in 
a dense medium created in Au+Au collisions.  It was proposed in~\cite{fries}
that the same processes contributing to ``jet quenching'' in the QGP
could produce photons (through Compton 
scattering or $q\overline{q}$ annihilation).  
These are also pQCD photons but will henceforth be referred to in the text 
as ``jet quenching'' photons, while those 
pQCD photons that are not produced due 
to the dense medium will be referred to as
``prompt'' photons.
Finally there are the coveted ``thermal'' photons,
($p_T \sim 1-4$~GeV), which are radiated from the Au+Au system and thus carry the temperature information of the system in its hottest
phases.

\section{Direct Photons at the SPS}

A measurement of direct photons has previously 
been made in heavy ion collisions at 
the CERN-SPS~\cite{WA98_photon}.  Experiment WA98 measured an excess 
of photons above those from hadronic sources of 10-20\% for 
$p_T > 1.5$~GeV/c in central Pb+Pb collisions.
The measured excess, a 1$\sigma$ effect, 
was found to be larger (by $\sim$ a factor of 2) 
than the expected direct photon signal from initial
pQCD processes, which was determined by measurements
of direct photons in p+Pb reactions at similar beam energies.
The measurements in p+Pb collisions and the scaling to Pb+Pb collisions 
have further uncertainties that must be taken into account when quantifying
this excess.  Strong conclusions cannot be made due to the experimental 
uncertainties on a relatively small direct photon signal 
above the large level of background.
However, the measurement of direct photons in Pb+Pb collisions has
provided constraints on the system temperature obtained from
calculations of photons radiated from
the QGP and/or hadron gas phases of a heavy ion collision at 
$\sqrt{s_{NN}} = 17$~GeV.  

\section{Results from PHENIX}

\begin{figure}[hbt]
\begin{center}
\mbox{\epsfxsize=6in\epsfbox{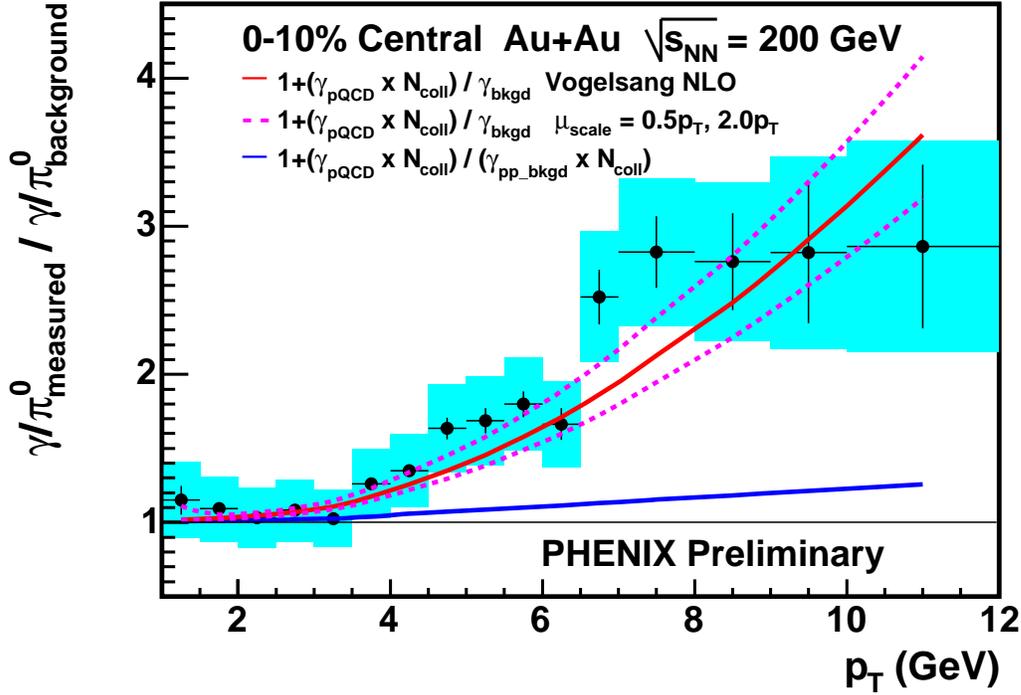}}
\end{center}
\caption{\label{fig:gam_cent}
Double ratio $(\gamma / \pi^0)_{measured}/(\gamma / \pi^0)_{background}$ for most central Au+Au collisions, 0-10\% centrality, indicating the fraction of direct photons (see Eq.~\ref{eq:ratio}).  The upper solid theory curve is 
$1 + \gamma_{theory}/\gamma_{background}$, where $\gamma_{theory}$ is a pQCD calculation~\cite{vogelsang} for p+p scaled by $N_{coll}$.  The dashed lines are the same, varying the mass scale in the pQCD calculation.  The lower solid theory curve is the expectation from pQCD if the $\pi^0$ were not suppressed (or the direct photons were equally suppressed), calculated by replacing $\gamma_{background}$ by
$\gamma_{p+p~background}~{\rm x}~N_{coll}$.}
\end{figure}

\begin{figure}[hbt]
\begin{center}
\mbox{\epsfxsize=6in\epsfbox{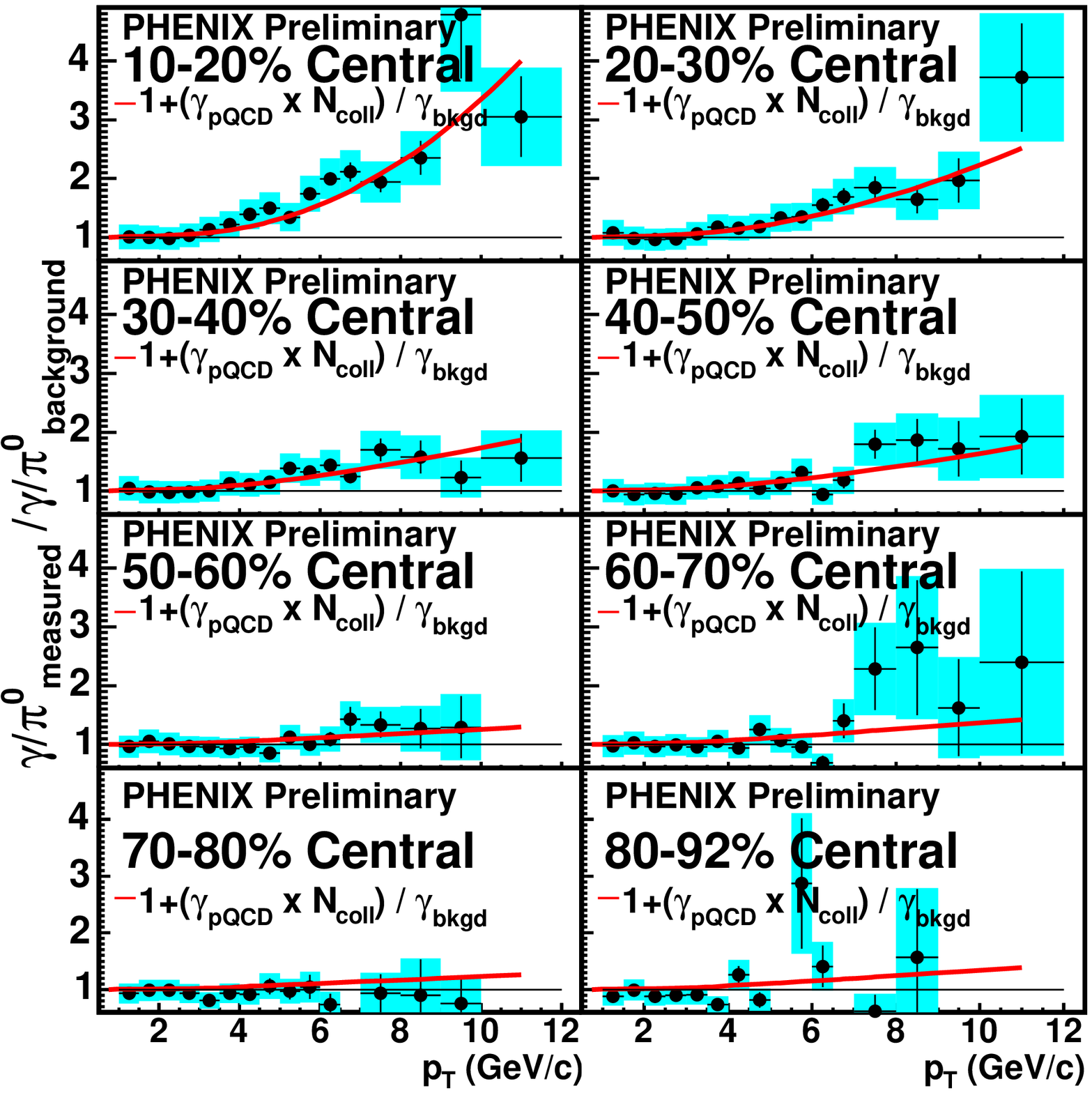}}
\end{center}
\caption{\label{fig:gam_allcent}
Double ratio $(\gamma / \pi^0)_{measured}/(\gamma / \pi^0)_{background}$ for Au+Au collisions, from central (top left) to peripheral (bottom right), indicating the fraction of direct photons (see Eq.~\ref{eq:ratio}).  The solid theory curve is $1 + \gamma_{theory}/\gamma_{background}$, where $\gamma_{theory}$ is a pQCD calculation~\cite{vogelsang} for p+p scaled by $N_{coll}$ for each centrality. }
\end{figure}

Direct photons are quantified as an excess above background,
expressed as the double ratio written in Eq.~\ref{eq:ratio}.  
\begin{eqnarray}
(\gamma / \pi^0)_{measured}/(\gamma / \pi^0)_{background} = \gamma_{measured}/\gamma_{background} \nonumber \\
~~ = (\gamma_{background} + \gamma_{direct})/\gamma_{background}
= 1 + \gamma_{direct}/\gamma_{background}
\label{eq:ratio}
\end{eqnarray}
The numerator $(\gamma / \pi^0)_{measured}$ is the point-by-point 
measurement of the photon yields divided by the $\pi^0$ yields, in which
much of the experimental uncertainties cancel.  The denominator
$(\gamma / \pi^0)_{background}$ is the simulated photon yields from 
hadronic decays, based on a fit to the measured $\pi^0$ spectrum, 
divided by the
fit.  The direct photon measurement is shown in terms of this ratio in Fig.~\ref{fig:gam_cent} for the most central Au+Au collisions and for all other centrality selections in Fig.~\ref{fig:gam_allcent} \cite{direct_photon}.
The PHENIX measurement in p+p collisions is not 
shown here (see~\cite{direct_photon}).
At high $p_T$, the direct photon yields are shown
to be consistent with pQCD calculations~\cite{vogelsang} 
and binary scaling ($N_{coll}$ scaling) for all centralities.  
In central Au+Au collisions, the direct photon content is 
found to be twice as large as the background photon content at 
$p_T \sim 10$~GeV, due to the large suppression of background 
photons in central collisions.
A large suppression of direct photons 
of factor 4-5, as seen in the $\pi^0$ yields 
at high $p_T$, can be ruled out. 
However, an interplay between a modest suppression
and increased production due to another mechanism in
heavy ion collisions, such as proposed in~\cite{fries}, cannot be ruled out.  
This will be further discussed in the following section.
A thermal photon signal, expected at $p_T \sim 1-4$~GeV/c cannot be extracted
from the current measurement because the systematic uncertainties 
are too large.  With the new Run-4 
data, we hope to be able to reduce the errors significantly. 

\section{Theoretical Calculations}

\begin{figure}[hbt]
\begin{center}
\mbox{\epsfxsize=6in\epsfbox{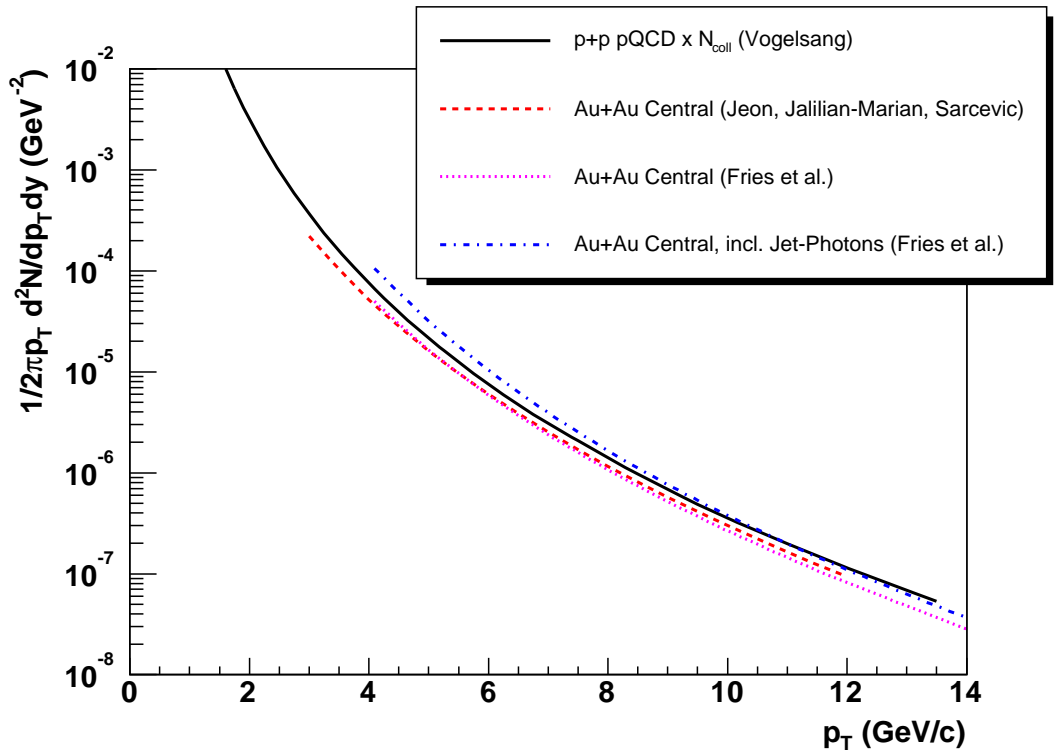}}
\end{center}
\caption{\label{fig:theory}Theoretical calculations of the photon yields as a function of $p_T$ in central Au+Au collisions.  No calculation of a thermal component is included in this plot.  See text for description of lines}
\end{figure}

In this section, theoretical calculations of direct photons, including
prompt photons and jet quenching photons, are compared.
Figure~\ref{fig:theory} shows the direct photon yields calculated for central 
Au+Au collisions.
The solid line is a pQCD calculation~\cite{vogelsang} for p+p collisions,
where the prompt photon yields are scaled by $N_{coll}$ for central Au+Au.
The dashed line is another calculation of only the prompt photons, but 
includes parton energy loss affecting the photons 
originating from bremsstrahlung processes (proposed in~\cite{jamal}).  
The dotted line is a different calculation including
this same effect~\cite{fries}, which agrees reasonably well with the former.
Finally the dashed-dotted line includes, in addition to this suppression, 
an enhancement
of direct photons due to the jet quenching production mechanism~\cite{fries}.  
A calculation of the thermal component is not included.
At $p_T = 4.25$~GeV/c, the dashed line, which includes 
parton energy loss in the medium, is $\sim 25$\% smaller than the solid line, which is pQCD with no medium modifications.  At the same $p_T$, the dashed-dotted line, which includes jet quenching photons, is $\sim 65$\% larger than the solid pQCD line.  The deviations from the $N_{coll}$-scaled pQCD calculation become smaller at higher $p_T$.  With reduced systematic errors, one will hopefully be able to distinguish between these scenarios. 

\section{Photon $v_2$}

\begin{figure}[hbt]
\begin{center}
\mbox{\epsfxsize=6in\epsfbox{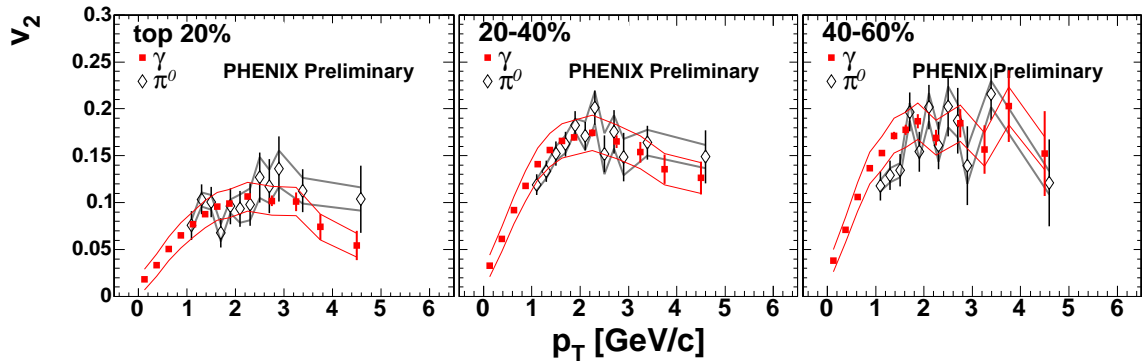}}
\end{center}
\caption{\label{fig:v2}Azimuthal anisotropy $v_2$ as a function of $p_T$ for inclusive, i.e. from all possible sources, photons (closed red squares) and $\pi^0$ (open black diamonds)~\cite{photon_v2}.}
\end{figure}

Another possible means to distinguish between jet quenching photons and 
the prompt photons is to measure the azimuthal asymmetry 
$v_2$ of the direct photons.
Since jet quenching is sensitive to the reaction plane, i.e. there is
more suppression out-of-plane than in-plane, we measure a positive $v_2$
for hadrons.  Correspondingly, the photons produced in this mechanism should
exhibit a negative $v_2$.  This, however, is another 
difficult measurement because
one has to disentangle the $v_2$ from the photons originating from 
hadronic decays, which is positive.  Figure~\ref{fig:v2} shows the measured $v_2$ for inclusive photons and for $\pi^0$.  
In the region of interest ($p_T \sim 4$~GeV/c), the errors are very large.  Again from Run 4, we hope to make this measurement with improved precision, and to higher $p_T$, in order to be able to deduce $v_2$ of the direct photons over a large $p_T$ range.

\section{Summary}

PHENIX has measured direct photons up to $p_T \sim 10$~GeV/c in centrality 
bins of 10\%.  
The yields of photons in excess of those from hadronic decays are consistent
with a pQCD calculation and binary scaling for all centralities.  
A large suppression
in central Au+Au collisions, as that observed for hadronic yields,
is ruled out by the data.  A modest suppression and/or enhancement cannot
be ruled out with the current systematic errors.  With the Run-4 data, we hope
to reduce the systematic uncertainties on the direct photon measurement to be
able to extract a thermal photon signal, if present.
To help distinguish between different sources of direct photons, we also plan to make a measurement of the $v_2$ of direct photons over a large $p_T$ range.
 
\section*{References}

\def\IJMPA{Int. J. Mod. Phys. A}
\def\JPG{J. Phys G}
\def\NCA{Nuovo Cimento}
\def\NIM{Nucl. Instrum. Methods}
\def\NIMA{Nucl. Instrum. Methods A}
\def\NPA{Nucl. Phys. A}
\def\NPB{Nucl. Phys. B}
\def\PLB{Phys. Lett. B}
\def\PLC{Phys. Repts.\ }
\def\PRL{Phys. Rev. Lett.\ }
\def\PRD{Phys. Rev. D}
\def\PRC{Phys. Rev. C}
\def\ZPC{Z. Phys. C}
\def\EPJC{Eur.Phys.J. C}

\end{document}